\documentclass{article}%

\usepackage{amsfonts}
\usepackage{amssymb}

\usepackage{amsmath, amsthm, graphicx,amssymb,wrapfig}

\newcommand{\cross}{\mbox{\boldmath $\times$}}

\newcommand{\bfL}{\mbox{\boldmath $L$}}

\def\aj{{AJ}}                   
\def\araa{{ARA\&A}}             
\def\apj{{ApJ}}                 
\def\aap{{A\&A}}                
\def\mnras{{MNRAS}}             

\def\ltsima{$\; \buildrel < \over \sim \;$}
\def\simlt{\lower.5ex\hbox{\ltsima}}
\def\gtsima{$\; \buildrel > \over \sim \;$}
\def\simgt{\lower.5ex\hbox{\gtsima}}
\def\gsim{ \lower .75ex \hbox{$\sim$} \llap{\raise .27ex \hbox{$>$}} }
\def\lsim{ \lower .75ex\hbox{$\sim$} \llap{\raise .27ex \hbox{$<$}} }
\def\gtrsim{ \lower .75ex \hbox{$\sim$} \llap{\raise .27ex \hbox{$>$}} }
\def\lesssim{ \lower .75ex\hbox{$\sim$} \llap{\raise .27ex \hbox{$<$}} }

\def\bfr{{\bf r}}

\def\half{{\textstyle{1\over2}}}
\def\bfr{{\bf r}}

\def\bfv{{\bf v}}

\def\bflambda{\mbox{\boldmath $\lambda$}}

\def\bfr{{\bf r}}
\def\bfx{{\bf x}}

\def\p{\partial}
\def\bfr{{\bf r}}

\def\bfv{{\bf v}}

\def\ffrac#1#2{{\textstyle\frac{#1}{#2}}}

\begin{document}

\title{A few of Michel H\'enon's contributions to dynamical astronomy}

\author{Scott Tremaine \\ Institute for Advanced Study \\ Princeton,
  NJ 08540, USA}

\maketitle

\begin{abstract}
This article reviews Michel H\'enon's contributions to a
  diverse set of problems in astrophysical dynamics, including violent
  relaxation, Saturn's rings, roundoff error in orbit integrations,
  and planet formation.
\end{abstract}

\noindent
I only met Michel H\'enon once, but I followed his tracks many
times. I first encountered his work in the early 1970s, when I was a
graduate student beginning to learn stellar dynamics. At that time the
subject was just beginning to blossom, and the selection of textbooks
and reviews was quite limited (\cite{cha42}, \cite{og65}, \cite{mr68}). Thus I was
delighted to discover MH's Saas-Fee lectures on ``Collisional dynamics
of spherical stellar systems'' \cite{mh73}, and I soon learned that
{\em any} paper by MH was worth reading, even if I had to use my very
limited French. It is sometimes said that papers by great physicists
such as Maxwell and Einstein are easier to understand than papers by
physicists who are good but not great, and MH's papers had a similar
quality, in that simple logic led to profound conclusions. Because
of space and time constraints, I cannot describe all of his work that
interested and impressed me, so I have tried to present a selection
that illustrates the unusual diversity of his contributions to
astrophysical dynamics, and focuses on topics in which we shared a
common interest.

\section*{Violent relaxation. I}

\noindent
The evolution of a cluster of stars or other self-gravitating N-body
system can crudely be divided into two phases. (i) On timescales short
compared to the relaxation time $t_{\rm relax}$ the dynamics is that
of a collisionless system, in which each star moves under the
influence of the smooth gravitational field generated by its $N-1$
siblings. The evolution is described by the collisionless Boltzmann
equation\footnote{Sometimes also called the Liouville equation, the
  Vlasov equation, or other names; but MH had strong opinions on the
  appropriate name to use \cite{hen82}, and I agree with him.} and the Poisson equation,
relating the phase-space density $f(\bfx,\bfv,t)$ and the
gravitational potential $\Phi(\bfx,t)$,
\begin{equation}
\frac{\p f}{\p t} + \bfv\cdot\frac{\p f}{\p \bfx} -
\frac{\p\Phi}{\p \bfx}\cdot\frac{\p f}{\p \bfv}=0, \qquad 
\nabla^2\Phi=4\pi G\int d\bfv f.
\end{equation}
A collisionless N-body system that is out of equilibrium will evolve
rapidly to an approximately stationary solution of the collisionless
Boltzmann equation. In cosmology this process is sometimes called
``virialization'' since the stationary system satisfies the
time-independent virial theorem $2K+W=0$, where $K$ and $W$ are the
kinetic and potential energy. Virialization is a rapid process: it
takes only a few times the crossing time $t_{\rm cross}=R/V$, where
$R$ and $V$ are a typical radius and velocity in the system. The
evolution is driven by large-scale oscillations in the gravitational
potential, which rapidly damp through phase mixing. (ii) On
timescales long compared to $t_{\rm relax}$ the evolution is driven by
two-body encounters between each star and its neighbors; the system
evolves slowly along a sequence of approximate solutions of the
collisionless Boltzmann equation; during this evolution the central density grows
while stars escape from the outer parts\footnote{The first substantive
discussion of this process was in MH's doctoral thesis at the
University of Paris \cite{hen61}.}.  The ratio of the relaxation and
crossing times is \cite{bt08}(e.g., eq.\ 1.38).
\begin{equation}
\frac{t_{\rm relax}}{t_{\rm cross}}\simeq \frac{0.1 N}{\ln N}.
\label{eq:trel}
\end{equation}

In the early 1960s N-body integrations of self-gravitating systems
were still extremely crude: a complete literature survey up to 1964,
so far as I know, consists of \cite{ho41}, \cite{vh60}, \cite{vh63},
and \cite{aar63}, all with $N\le100$. The results from simulations of
such small systems are difficult to interpret, not just because of
statistical fluctuations but also because equation (\ref{eq:trel})
implies that the relaxation and crossing times are not well-separated
unless $N\gg 100$, so the two phases of evolution are not distinct.

MH's elegant idea \cite{hen64}, was to restrict the N-body system to
spherical symmetry, replacing the point masses by spherical
shells (the idea is originally presented by \cite{cam62}, who in turn
attributes it to George Gamow). Shell $j$ has mass $m_j$, radius
$r_j(t)$, and conserved angular momentum per unit mass $L_j$. The
equations of motion are
\begin{equation}
\frac{d^2r_j}{dt^2}=\frac{L_j^2}{r_j^3}-\frac{GM_j}{r_j^2}, \qquad
M_j=\sum_{r_k<r_j}m_k + \ffrac{1}{2} m_j.
\end{equation}
As a result of this simple ansatz, (i) two-body relaxation is greatly reduced,
because the force field from a spherical shell is much smoother than
the force field from a point; (ii) the equations of motion are
simpler; for example one follows one coordinate instead of three and
the equations can be solved analytically between shell crossings; (iii)
the simulation preserves the spherical symmetry that one expects to
see in large-N systems.

MH followed several systems with $N\le 100$, starting from ``top hat''
initial conditions (uniform density inside a spherical volume), with a
Maxwellian velocity distribution and initial virial ratio
$-2K/W=0.5$. A later calculation with $N=1000$ is described in
\cite{hen68}. These crude simulations yielded the first clear
description of the process of virialization. MH showed that the final
stationary states depended somewhat on the initial conditions but all
shared certain properties, including the presence of a dense central
core, an extended envelope in which the density decayed as
$\rho(r)\sim r^{-4}$, and a velocity-dispersion tensor that was mainly
radial in the envelope. This work was the precursor to Lynden-Bell's
(1967) seminal work on virialization (which he termed ``violent
relaxation''), and to the ``basis-function'' or ``self-consistent
field'' N-body codes \cite{vil82,ho92,saha93} that are now widely
used to study galaxy mergers, galaxy stability \cite{hen73}, and
other collisionless collective processes.

\section*{Violent relaxation. II}

\noindent
In the mid-1980s I became interested in whether arguments from
statistical mechanics could be used to predict the stationary solutions
of the collisionless Boltzmann equation that resulted from violent
relaxation. Typically, attempts to do this , e.g.{lb67}, lead
to solutions with infinite mass, so my goal was to address a more
modest question: if $f_i(\bfx,\bfv)$ and $f_f(\bfx,\bfv)$ are the
phase-space distribution functions of the stellar system before and
after violent relaxation, what constraints can be placed on $f_f$
given $f_i$, other than mass, energy, and angular-momentum conservation?

Violent relaxation can be thought of as a Markov process in which mass
elements are shuffled among cells in phase space. By definition, any
Liapunov functional $L[f]$ of the Markov process decreases at each
step of the process; thus, if $L[f_f]\le L[f_i]$ for all Liapunov
functionals, then $f_f$ is said to be more mixed than $f_i$
(e.g. \cite{weh78,gor10}). I was able to find a simple criterion
for this partial ordering---more accurately, I was able to look it up
in \cite{lhp52}---which was later simplified further by
\cite{deh05}: define a one-parameter family of functionals
\begin{equation}
D_\phi [f]=\int d\bfx d\bfv \mbox{ max\,}[f(\bfx,\bfv)-\phi,0].
\label{eq:mh}
\end{equation}
Then $f_f$ is more mixed than $f_i$ if and only if $D_\phi[f_f] \le
D_\phi[f_i]$ for all $\phi$. One corollary of this approach is that
the standard Boltzmann entropy $S=-\int d\bfx d\bfv \,f\log f$ plays no
special role in violent relaxation, since $-S$ is only one of many
possible Liapunov functionals of the form $\int d\bfx d\bfv \,C(f)$
where $C$ is a convex function.

While preparing that work for publication, I learned to my surprise
that MH and Donald Lynden--Bell had worked on the same problem two
decades earlier, without publishing their results. I got copies of
their notes and correspondence and discovered that they had obtained
almost all the same results that I had---although their achievement
was much more impressive since most of the work in the statistical
mechanics literature that I used had not been published at that
time. Moreover, MH had the following beautiful physical analogy to
explain the result (\ref{eq:mh}): Consider a hill in which the height
above sea-level is $h_i(x,y)$. An engineer wishes to transform the
hill to a differently shaped one, with a new topography $h_f(x,y)$. We
assume that there is only one summit or local maximum in both
$h_i(x,y)$ and $h_f(x,y)$. The engineer has a bulldozer that can
scrape off earth from the hill and push it downhill; however, the
bulldozer is not powerful enough to push the material uphill. It is
not hard to see that the engineer can succeed at this task if and only
if there is more earth above every height $H$ in the initial hill than
in the final hill, that is, if $D_H[h_f]\le D_H[h_i]$, which is
precisely analogous to condition (\ref{eq:mh}).

After some correspondence the three of us agreed to publish the paper
together \cite{thlb86}; not only was I proud to be able to
collaborate with two of my scientific heroes, but in the course of our
discussions MH suggested that we try a new communication medium called
e-mail and sent me my first e-mail message!

I think the main impact of this work has been to dampen the enthusiasm
of theorists for explaining ``universal'' profiles resulting from
virialization or violent relaxation, such as the famous NFW profile
\cite{nfw97}, in terms of maximum-entropy arguments: such arguments
cannot give a unique final state unless one first shows why some
particular entropy $-\int d\bfx d\bfv C(f)$  is the only relevant one.

\section*{Roundoff error}

\noindent
Roundoff error has always plagued numerical explorations of dynamical
systems, in particular long integrations of planetary systems (e.g. \cite{hay08}). The
problem with roundoff error is not so much that phase-space positions and
other parameters cannot be represented exactly, but rather that
it accumulates with time; in a prescient paper
\cite{new99} pointed out that roundoff errors in planetary positions
should grow with time at $t^{3/2}$. In fact, Newcomb was
over-optimistic: unless one is very careful the errors in modern
computations tend to grow at $t^2$. The growth of roundoff error is
inevitable in floating-point arithmetic because floating-point
operations are many-to-one maps rather than one-to-one maps. As
computers become faster and integration algorithms became more accurate, roundoff
error has played a growing role in the error budget of studies of the
long-term behavior of dynamical systems.  

Perhaps the most important advance in controlling roundoff error was
the introduction in 1985 of the IEEE Standard for floating-point
arithmetic (IEEE 754), which has now been adopted (more or less) in
most compilers. Among its features, IEEE 754 requires that all
arithmetic operations are rounded to the nearest number that can be
represented exactly in the computer, and in the case of ties rounded
to the nearest number whose least significant bit is even. IEEE 754
helps us to manage the disease of roundoff error but does not cure it.

MH studied this problem with his student Fran\c oise Rannou
\cite{ran74}. First, they reduced the problem to its simplest form by
studying an area-preserving map in a compact phase space of two
dimensions rather than a Hamiltonian system, a technique made famous a
few years early by MH's work with another student, Carl Heiles
\cite{hh64}. The map was chosen to exhibit both invariant curves and
chaotic regions. They then modified the map slightly so it was a
one-to-one map of integers onto integers---in other words, they
changed the map into a permutation of points on a lattice---and
studied the properties of the modified map using fixed-point
arithmetic, which has no roundoff error. An additional advantage of
the lattice map is that all orbits are periodic so the properties of
the map can be characterized completely with finite computing
resources.

The most striking of Rannou's conclusions was that the lattice maps
bear a strong visual resemblance to the original floating-point maps,
with invariant curves in the original map corresponding to periodic
orbits with short cycles and chaotic orbits to long cycles. This and
other findings, in her understated conclusions, ``encourage the view
that conventional computer studies are not seriously affected by
roundoff errors''.

In the early 1990s I worked on lattice maps for Hamiltonian systems
with David Earn, who was then a student at Toronto. Partway
through the work, we were disappointed to learn---although by this
time I should not have been surprised---that MH and Rannou had covered much of
the ground long before us. We extended Rannou's work with more
thorough numerical calculations, and were able to prove that
symplectic maps from $R^{2N}$ to itself could be replaced by 
maps that were the restriction to a lattice of a symplectic map that
was ``close'' to the original map in a precise sense \cite{et92};
thus lattice maps can be thought of as replacing forward error
analysis with backward error analysis. The lattice maps differ from
the original symplectic map by a small but rapidly varying
perturbation; as the lattice spacing shrinks the difference becomes
smaller but it varies more and more rapidly. Thus the
Kolmogorov-Arnold-Moser or KAM theorem does not guarantee that
invariant tori in the original map will survive in the lattice map, no
matter how small the spacing, although in practice the tori seemed to
be quite robust. The question of whether integer or floating-point
calculations are more faithful representations of the long-term
behavior of Hamiltonian systems remains open.

Perhaps the simplest demonstration of the long-term effects of roundoff
error comes from the rotation map
\begin{equation}
x_{n+1}=x_n\cos\theta - y_n\sin\theta, \quad y_{n+1}=x_n\sin\theta +
y_n\cos\theta,
\end{equation}
with fixed angle $\theta$ and initial conditions $x_0=1$, $y_0=0$. The
map should conserve $E_n\equiv x_n^2+y_n^2$ but numerical computations
generally show a linear drift, $E_n-1 \propto n$, with the slope of
the drift depending erratically on the rotation angle $\theta$. Thomas
Quinn and I investigated this map as a simple model for the growth of
roundoff error in long orbit integrations \cite{qt90}, and we found
recipes to reduce the drift to $E_n-1\propto n^{1/2}$. A central
ingredient in these recipes is to correct for the fact that
$\cos^2\theta+\sin^2\theta-1$ is typically not exactly zero when
evaluated with floating-point arithmetic.

\cite{hp98} looked at this problem more deeply: in floating-point
arithmetic, finding a value of $\theta$ for which $\cos^2\theta+\sin^2\theta-1$ is
close to zero is equivalent to the Diophantine problem of finding
integers $m$ and $n$ such that $m^2+n^2=2^{2p}+k$ where $p=53$ for
IEEE 754 double-precision arithmetic and $|k|$ is as small as
possible. They were able to show that the best solutions have $k=1$,
although there are only eight for $0\le\theta\le\pi/4$, and to give
explicit algorithms for finding the many solutions for which $|k|$
exceeds unity but is still small. 

Regrettably, MH's work on roundoff error has not received much
attention. Most computational astrophysicists still adopt the
convenient belief that ``double precision is accurate enough''. Often
it is, but not always. Like many other arcane problems in theoretical
computer science, the problems studied by MH in this subject may prove to
be important only many years after they were first posed and
solved.

\section*{Apples in a spacecraft}

\noindent
This controversy began with a thought experiment by the Nobel Prize
winner Hannes \cite{alf71}. Consider a spacecraft in a circular
orbit around the Earth. The spacecraft is assumed to be synchronously
rotating (the same side always faces the Earth). Now release a swarm
of inelastically colliding objects (``apples'') in the spacecraft;
what will be their final state? Alfv\'en concluded that the apples
would collect in a pile at the center of mass of the spacecraft.

The following argument is equivalent to Alfv\'en's, but simpler and
more concise.  Let $m_i$, $\bfr_i$, and $\bfv_i$, $i=0,\ldots,N$ be
the masses, positions, and velocities of the spacecraft ($i=0$) and
the $N$ apples, in an inertial frame. The energy and angular momentum
of this collection are
\begin{equation}
E=\sum_{i=0}^N m_i\Big(\half v_i^2 - \frac{GM_\oplus}{r_i}\Big), \quad \bfL=\sum_{i=0}^N m_i\,\bfr_i\cross\bfv_i.
\end{equation}
We extremize the energy at fixed angular momentum. Using a vector Lagrange
multiplier $\bflambda$, the condition for an extremum is
\begin{equation}
0=\delta E - \bflambda\cdot\delta\bfL=\sum_{i=0}^N
m_i\left[\delta\bfv_i\cdot(\bfv_i-\bflambda\cross\bfr_i)
  +\delta\bfr_i\cdot\Big(\frac{GM_\oplus}{r_i^3}\bfr_i+\bflambda\cross\bfv_i\Big)\right].
\end{equation}
This requires
\begin{equation}
\bfv_i=\bflambda\cross\bfr_i, \quad \lambda^2=\frac{GM_\oplus}{r_i^3},
\quad \bflambda\cdot\bfr_i=0;
\label{eq:alfven}
\end{equation}
in words, all of the orbits must be coplanar and circular, with the
same orbital radius. 

The importance of Alfv\'en's thought experiment is that it suggests a
way to collect small bodies such as planetesimals into dense
agglomerations that might form planets (these were called ``jet
streams'' by their proponents). 

Some years later, \cite{hen78} wrote a short but crushing reply,
which began ``Unfortunately, Alfv\'en's reasoning is incorrect and the
final state of the system is in reality rather different from what he
predicts. This conclusion has apparently escaped notice so far, and
Alfv\'en's result continues to be cited uncritically. Thus it appears
desirable to correct the record.'' Although MH's reasoning was
physical rather than mathematical---there are no equations in his paper, compared to two dozen
in Alfv\'en's---it is simplest to describe in the context of the
derivation above. The method of Lagrange multipliers finds an extremum
in the energy at fixed angular momentum, but the extremum represented
by equation (\ref{eq:alfven}) is a saddle point, not a minimum. The
minimum energy state occurs when half the apples are on the floor of
the spacecraft (the point closest to Earth) and the other half on the
ceiling. Dissipation acts to disperse the radii of the apples, not to
bring them together; Alfv\'en's model is an argument against jet
streams rather than in favor of them, and indeed this concept no
longer plays a role in models of planet formation\footnote{In the early 1970s I was
  a graduate student in physics at Princeton, which then had a series
  of written general exams that had to be completed before starting a
  thesis. One of the problems on the exam in spring 1971 was to prove
  Alfv\'en's result. I'm proud to report that my fellow students, like
  MH, recognized that the result was wrong, although only after the
  exam, and took pleasure in pointing this out to the professors.}. 

\section*{Saturn's rings}

\noindent
When NASA's Voyager 1 and 2 spacecraft flew past Saturn in 1980 and
1981, they revealed that the planet's famous rings were far more
complex than previously suspected (see \cite{gt82} for a
post-Voyager review and \cite{cuz10} for a recent one). Among the
puzzles emerging from the spacecraft data were the following: (i)
Collisions between ring particles redistribute angular momentum so the
ring spreads, much like the gaseous accretion disks in other
astrophysical systems, and sharp radial features are washed out.
However, Voyager revealed a rich spectrum of narrow gaps, ringlets,
and other features on all scales down to the spacecraft resolution
limit of $\sim10\mbox{\,km}$. (ii) The thickness of the rings can be
estimated by photometric observations during the Earth's passages
through the ring plane, which occur every 13 years. These suggest that
the ring thickness is $H\sim 1\mbox{\,km}$. However, inelastic
collisions between the ring particles rapidly damp any motions normal
to the ring plane, so the ring should be much thinner than the
ring-plane-crossing observations imply.


MH's work on Saturn's rings began with a simple question: {\em is} there
a typical size of a ring particle \cite{hen81,hen84}? Natural processes
such as grinding, fragmentation by high-velocity impacts, or
coagulation generally produce power-law distributions of particle size
over several orders of magnitude, and such size distributions are seen
in asteroids, meteorites, and debris on the lunar surface.  Moreover the
exponents of such power laws have a relatively narrow range: if the
number of particles in a small radius range is
\begin{equation}
dN= \left(\frac{r_0}{r}\right)^\beta\,d\log r
\label{eq:pow}
\end{equation}
then $\beta\simeq 1.8$--$3.6$ (e.g. \cite{har69}).  Thus the ring
particles may be better characterized by a power-law distribution
than by a single size.

MH explored the implications of this ansatz. In the following
discussion, we take $dN$ in equation (\ref{eq:pow}) to refer to the
total number of ring particles in the radius range $d\log r$, and
assume that this power law applies for all radii larger than $r_{\rm
  min}$, which is several orders of magnitude smaller than $r_0$. With
this normalization $r_0$ is approximately the size of the largest
particle in the rings.

If the ring is flat and the centers of the particles lie in the ring's
midplane, the apparent thickness $H$ of the edge-on ring is given
approximately by the diameter of the largest particles for which the
optical depth of the edge-on ring is unity. This yields
\begin{equation}
H\left(\frac{r_0}{H}\right)^\beta\simeq R
\label{eq:one}
\end{equation}
where $R\simeq 10^{10}\mbox{\,cm}$ is the radius of the rings. 

The ring particles can only efficiently scatter radiation of
wavelength $\lambda$ if $r\gtrsim \lambda/(2\pi)$.  Pre-Voyager
ground-based measurements of the radar cross-section and radio
brightness temperature of the rings could therefore be used to
constrain the distribution of ring particles on size scales of a few
cm. They implied that the normal geometrical optical depth of the
rings is of order unity for $r=r_r\simeq 4\mbox{\,cm}$, or
\begin{equation}
r_r^2\left(\frac{r_0}{r_r}\right)^\beta\simeq 2R\Delta R
\label{eq:two}
\end{equation}
where $\Delta R\simeq 4\times10^9\mbox{\,cm}$ is the radial range of
the densest part of the rings.  Equations (\ref{eq:one}) and
(\ref{eq:two}) are sufficient to determine the parameters of the size
distribution\footnote{The numbers given here can differ from the numbers given by
MH by up to a factor of two or so. These differences are not important.}:
\begin{equation}
\beta=3.1, \quad r_0=40\mbox{\, km}.
\end{equation}

The ring particles orbit inside their Roche limit, given by 
\begin{equation}
R_{\rm
  Roche}=1.523(M/\rho)^{1/3}=1.31\times10^{10}\mbox{\,cm}\left(\frac{0.9\mbox{\,g
      cm}^{-3}}{\rho}\right)^{1/3},
\label{eq:roche}
\end{equation}
where $M=5.684\times10^{29}\mbox{\,g}$ is the mass of Saturn and the
particle density $\rho$ is given relative to that of ice. The Roche limit
is the orbital radius at which a fluid satellite will be tidally
disrupted icy objects as large as $r_0$ can be
held together in the tidal field of Saturn by their internal strength.

Since the exponent $\beta$ is larger than 3, the ring mass
is dominated by small particles rather than large ones. We have 
\begin{equation}
M_{\rm ring}= \ffrac{4}{3}\pi \rho \int  d\log r\, r^3
\left(\frac{r_0}{r}\right)^\beta=\frac{4\pi}{3(\beta-3)}
\rho \,r_0^\beta r_{\rm min}^{3-\beta}=1.1\times
10^{22}\mbox{\,g}\,\frac{\rho}{0.9\mbox{\, g cm}^{-3}}\left(\frac{1\mbox{\,cm}}{r_{\rm min}}\right)^{0.1}.
\label{eq:mring}
\end{equation}
The result depends only weakly on the poorly known minimum size
$r_{\rm min}$. This estimate of the ring mass is consistent with an
independent estimate from the dispersion relation for density waves,
$M_{\rm ring}=(3\pm 2)\times 10^{22}\mbox{\,g}$ \cite{esp83},
although this may be an underestimate because density waves are not
present in the densest parts of the rings.

The wide range of particle sizes in the rings has other
consequences. Massive particles traveling on circular orbits
gravitationally repel particles on nearby orbits: at each conjunction
the gravitational force from the massive particle excites radial
oscillations (eccentricity) in the smaller particle, and because the
Jacobi constant is conserved the enhanced eccentricity results in a
transfer of angular momentum from the inner particle to the outer
one. The orbit-averaged gravitational torque between two masses $m_1$
and $m_2\ll m_1$ on circular orbits of radii $R$ and $R+d$, $d\ll r$,
is \cite{gt80}
\begin{equation}
T=  0.399\frac{Gm_2^2m_1R^3}{M d^4}
\label{eq:tt}
\end{equation}
where $M$ is the mass of the central body. This
expression is only valid if $d\gtrsim r_H$ where
$r_H=R[(m_1+m_2)/3M]^{1/3}\simeq R(m_1/3M)^{1/3}$ is the mutual Hill radius of the two
particles; for smaller separations a rough approximation to the torque
is obtained by replacing $d$ by $r_H$ in equation (\ref{eq:tt}). 

To analyze the effects of these torques, MH made a crude division of
the ring particles, at a radius $r_g$ to be determined below, into
``big'' and ``small'' particles. Big particles are massive enough that
they can overcome the spreading of the rings due to collisions and
open up a gap around themselves, while small particles cannot. We can now
estimate the integrated angular-momentum current from the
gravitational torques between the small particles. Using equation
(\ref{eq:pow}) for the number $dN$ of ring particles in a small size
range, we can write the number of particles in a small range of size
and orbital radius as $dN dR/\Delta R$ where $\Delta R$ is the ring
width. The mass of a particle is $\frac{4}{3}\pi\rho r^3$. Since the
torque (\ref{eq:tt}) falls off rapidly with separation $d$ we can
approximate it as zero for $d\gtrsim r_H$ and equal to $\sim
Gm_2^2m_1R^3/(Mr_H^4)$ for $d\lesssim r_H$. Then the angular-momentum
current is equal to the torque exerted by small particles inside $R$
on small particles outside $R$, or
\begin{equation}
C_L^{\rm ring}\simeq \frac{GR^3}{M\Delta R^2}\int_0^{r_g} dN(r)\int_0^r
dN(r') \frac{m(r)^2m(r')}{r_H^2(r)}\simeq \frac{GR\rho^{4/3}}{
M^{1/3}\Delta R^2}M_{\rm ring}r_0^\beta r_g^{4-\beta};
\end{equation}
here $M_{\rm ring}$ is the ring mass (eq.\ \ref{eq:mring}) and we have
assumed $\beta>3$. Since the ring is close to the Roche limit (eq.\
\ref{eq:roche}) $\rho\simeq M/R^3$ so this result simplifies to 
\begin{equation}
C_L^{\rm ring}\simeq \frac{G\rho}{\Delta
R^2}M_{\rm ring}r_0^\beta r_g^{4-\beta}.
\end{equation}

A similar calculation yields the torque produced by a ``big'' particle 
of mass $m$ on a uniform ring separated from it by a gap $d\gtrsim r_H$:
\begin{equation}
C_L^{\rm gap}\simeq \frac{Gm^2 R^3}{M\Delta R d^3}M_{\rm ring}.
\end{equation}
This is the angular-momentum current across the gap, which in a steady state
must equal the current $C_L^{\rm ring}$ through the ring on either side of
the gap, so we find
\begin{equation}
d^3(r)\simeq \frac{m^2R^3\Delta R\, r_g^{\beta-4}}{\rho M r_0^\beta} \simeq
\frac{r^6 r_g^{\beta-4}\Delta R}{r_0^\beta};
\label{eq:gap}
\end{equation}
in the last expression we have again set $\rho\simeq M/R^3$.  The
division between big and small particles is at the radius $r_g$ where
$d\sim r_H$, which yields $r_g\simeq (r_0^\beta/\Delta
R)^{1/(\beta-1)}\simeq 1.5\mbox{\,km}$. The minimum gap size is $\sim
2r_g$ or a few kilometers, and the gap size scales with the size of
the big particle as $d\propto r^2$. The number of big particles, and
thus the number of gaps, is given by equation (\ref{eq:pow}) as
$N_{\rm gap}=\simeq \int_{r_g}^{r_0} dN\simeq 8\times 10^3$. The total
width of the gaps is
\begin{equation}
\Delta R_{\rm gap} \simeq \int_{r_g}^{r_0} dN d(r)\simeq \Delta R.
\end{equation}
Remarkably, this result is independent of the exponent $\beta$ in the
size distribution; it implies that, in MH's words, the ring ``settles
automatically into a state in which the gaps and the ringlets occupy
comparable areas''---the geometrical optical depth is either zero (in
the gaps) or much larger than unity (in the ringlets between the
gaps), but when averaged over scales larger than the typical gap size,
as in most observations, the optical depth is of order unity.

\begin{figure}[ht]
\includegraphics[width=0.5\textwidth]{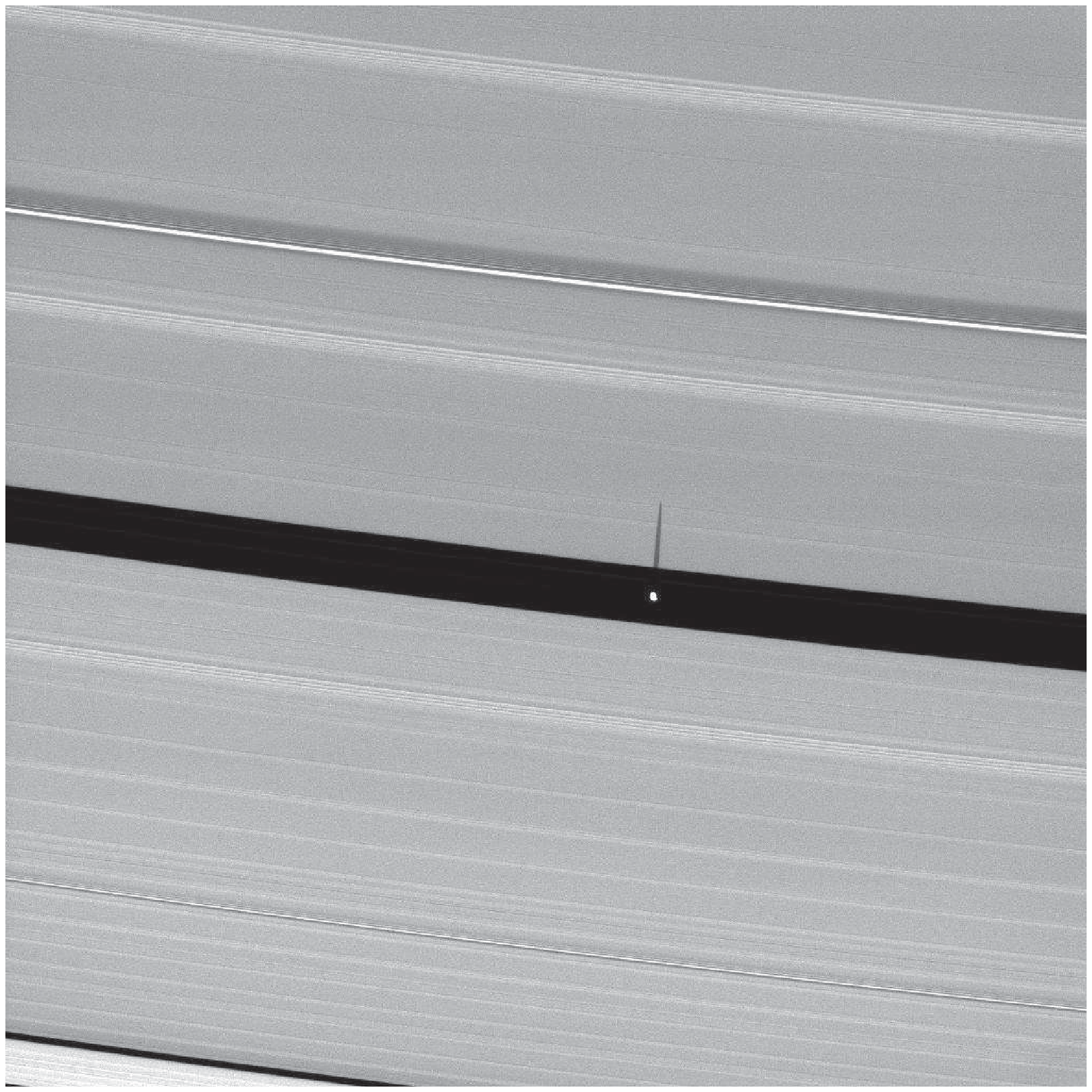}
\includegraphics[width=0.5\textwidth]{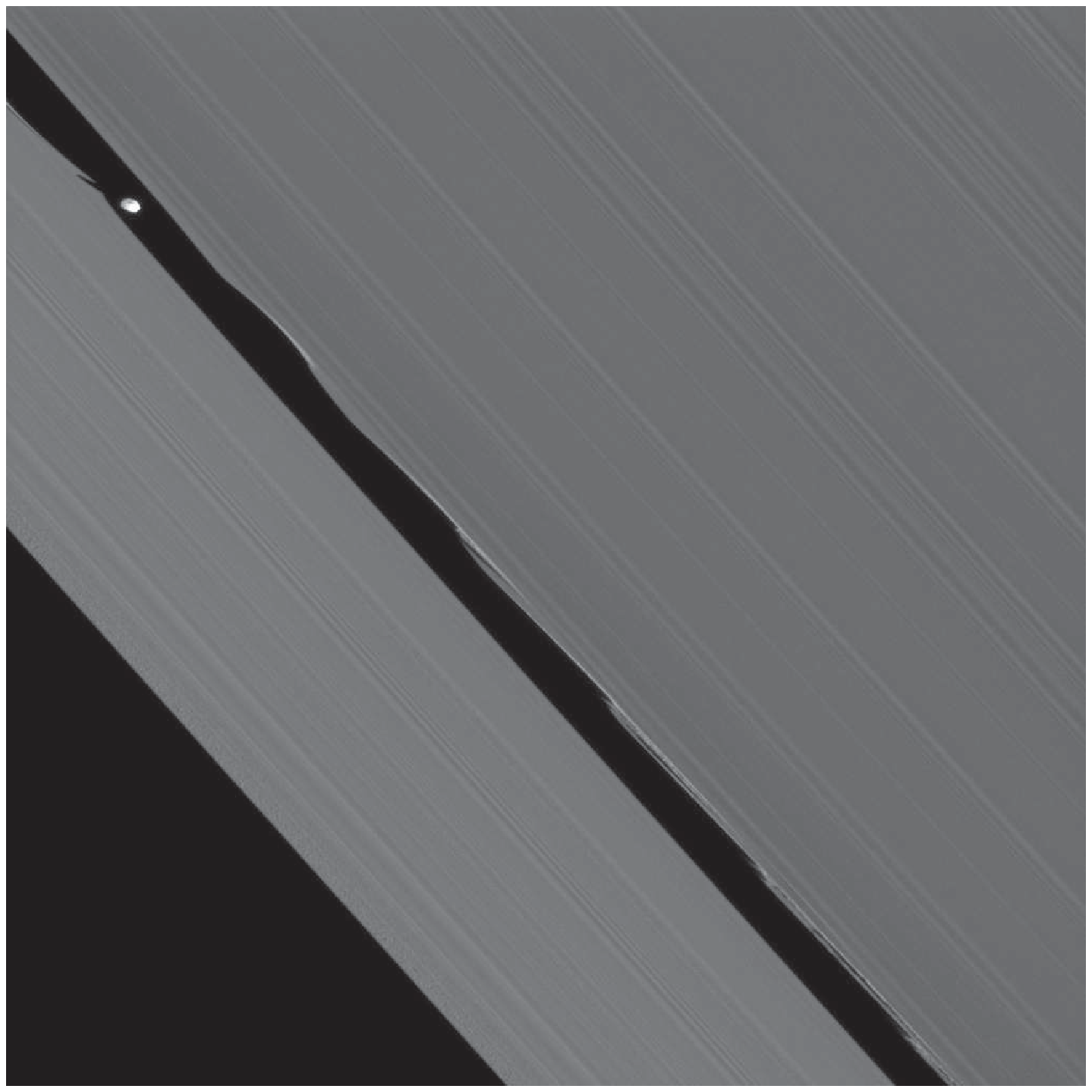}
\caption{\small (left) The satellite Pan, which orbits inside the 325-km wide
  Encke gap in Saturn's rings. The dark streak is Pan's shadow, which
  is long because the Sun is close to the ring plane. (right) The
  satellite Daphnis, which orbits inside the 42-km wide Keeler
  gap. The periodic structures at the edges of the gap are bending
  waves about 1 km high, excited because Daphnis has a small orbital
  inclination ($0.004^\circ$) relative to the rings.}
\label{fig:one}
\end{figure}

Thus one simple assumption, that the distribution of particle sizes in
the rings is a power law, led to a rich and detailed model that
explained many of the then-known features of Saturn's rings.  Sadly,
subsequent observations, particularly by the Cassini spacecraft since
it arrived in Saturn orbit in 2004, were an example of what Thomas Huxley
called ``The great tragedy of science---the slaying of a beautiful
hypothesis by an ugly fact''. The first problem to emerge is that the
rings are not flat: in particular the satellite Mimas, which is on an
orbit with an inclination of $1.57^\circ$ (relative to Saturn's
equator, which is also the ring plane), excites bending waves at the
5:3 orbital resonance in the rings. The amplitude of these waves is
$\sim 0.5\mbox{\,km}$, large enough to explain the apparent thickness
$H$ of the edge-on ring \cite{shu83}. The actual ring thickness is
probably no more than a few tens of meters. A second problem is that
thorough searches by Cassini have discovered only two satellites
orbiting in gaps within the rings: Pan and Daphnis, with radii of
$14\mbox{\,km}$ and $4\mbox{\,km}$, compared to MH's estimate of
almost $10^4$ satellites larger than $r_g\simeq 1.5\mbox{\,km}$ (see Figure). There
is also a handful of known satellites smaller than $r_g$ that have not
cleared gaps: these include S/2009 S1, with a radius of about
$0.15\mbox{\,km}$, and ``propellor moonlets'', which produce S-shaped
wakes a few km long. Third, observations of stellar occultations by
the rings (both from the ground and spacecraft) and spacecraft radio
transmissions through the rings imply a particle size distribution
with index $\beta\simeq 2$ and upper cutoff $r_{\rm max}\sim
10$--$20\mbox{\,m}$, varying significantly with radial distance from
Saturn \cite{cuz09}. Above this radius, the distribution of sizes
steepens sharply, consistent with the small number of propellor
moonlets and gap-opening satellites.

The processes that determine the distribution of sizes of Saturn's
ring particles are not understood, but apparently they do not lead to
the simple power-law distribution over more than five orders of
magnitude that led MH to his elegant model.

\section*{Final remarks}

\noindent
In preparing this retrospective, I have been forced to leave out
several of my favorites among MH's contributions: his analysis of the
isochrone potential, whose simple analytic properties could have been
discovered by Newton; his numerical simulations of globular cluster
evolution; ``H\'enon's paradox'' on the escape of stars from clusters;
the H\'enon--Heiles potential and his work on the third integral in
galactic dynamics; his comprehensive numerical study of the restricted
three-body problem, which laid to rest many questions that had been
unanswered for centuries; and his unexpected analytic solution of the
dynamics of the Toda lattice. I hope I have captured enough of the
beauty of Michel H\'enon's research that some readers will
take the time to appreciate these other contributions as well.

\end{document}